\pdfoutput=1
\RequirePackage{ifpdf}
\ifpdf 
\documentclass[pdftex]{sigma}
\else
\documentclass{sigma}
\fi

\renewcommand{\leq}{\leqslant}
\renewcommand{\geq}{\geqslant}

\renewcommand{\Im}{\operatorname{Im}}
\newcommand{\tr}{\operatorname{tr}}

\newcommand{\Ai}{\operatorname{Ai}}
\newcommand{\Bi}{\operatorname{Bi}}

\newcommand{\sgn}{\operatorname{sign}}
\makeatother

\newcommand{\C}{{\mathbb C}}

\newcommand{\R}{{\mathbb R}}

\newcommand{\1}{{\chi}}
\newcommand{\prob}{{\mathbb P}}

\newcommand{\conv}[1]{\overset{#1}{\longrightarrow}}

\numberwithin{equation}{section}

\newtheorem{Theorem}{Theorem}[section]
\newtheorem{Lemma}[Theorem]{Lemma}
 { \theoremstyle{definition}
\newtheorem{Example}[Theorem]{Example}
\newtheorem{Remark}[Theorem]{Remark} }

\begin{document}

\allowdisplaybreaks

\renewcommand{\thefootnote}{$\star$}

\newcommand{\arXivNumber}{1104.0153}

\renewcommand{\PaperNumber}{083}

\FirstPageHeading

\ShortArticleName{On the Scaling Limits of Determinantal Point Processes}

\ArticleName{On the Scaling Limits of Determinantal\\ Point Processes with Kernels Induced\\ by Sturm--Liouville Operators\footnote{This paper is a~contribution to the Special Issue on Asymptotics and Universality in Random Matrices, Random Growth Processes, Integrable Systems and Statistical Physics in honor of Percy Deift and Craig Tracy. The full collection is available at \href{http://www.emis.de/journals/SIGMA/Deift-Tracy.html}{http://www.emis.de/journals/SIGMA/Deift-Tracy.html}}}

\Author{Folkmar BORNEMANN}

\AuthorNameForHeading{F.~Bornemann}

\Address{Zentrum Mathematik -- M3, Technische Universit\"at M\"unchen, 80290~M\"unchen, Germany}
\Email{\href{mailto:bornemann@tum.de}{bornemann@tum.de}}
\URLaddress{\url{http://www-m3.ma.tum.de/bornemann}}

\ArticleDates{Received April 15, 2016, in f\/inal form August 16, 2016; Published online August 19, 2016}	

\Abstract{By applying an idea of Borodin and Olshanski [\textit{J.~Algebra} \textbf{313} (2007), 40--60], we study various scaling limits of determinantal point processes with trace class projection kernels given by spectral projections of selfadjoint Sturm--Liouville operators. Instead of studying the convergence of the kernels as functions, the method directly addresses the strong convergence of the induced integral operators. We show that, for this notion of convergence, the Dyson, Airy, and Bessel kernels are universal in the bulk, soft-edge, and hard-edge scaling limits. This result allows us to give a short and unif\/ied derivation of the known formulae for the scaling limits of the classical random matrix ensembles with unitary invariance, that is, the Gaussian unitary ensemble (GUE), the Wishart or Laguerre unitary ensemble (LUE), and the MANOVA (multivariate analysis of variance) or Jacobi unitary ensemble (JUE).}


\Classification{15B52; 34B24; 33C45}

\begin{flushright}
{\em Dedicated to Percy Deift at the occasion of his 70th birthday.}
\end{flushright}

\renewcommand{\thefootnote}{\arabic{footnote}}
\setcounter{footnote}{0}

\section{Introduction}\label{sect:intro}

We consider determinantal point processes on a (not necessarily bounded) interval $\Lambda = (a,b)$ with a~correlation kernel given by a trace class projection kernel,
\begin{gather}\label{eq:prokernel}
K_n(x,y) = \sum_{j=0}^{n-1} \phi_{j}(x) \phi_{j}(y),
\end{gather}
where $\phi_{0},\phi_1,\ldots,\phi_{n-1}$ are orthonormal in $L^2(\Lambda)$; each $\phi_j$ may have some dependence on $n$ that we suppress from the notation.
We recall (see, e.g., \cite[Section~4.2]{AGZ}) that for such processes the joint probability density of the $n$ points is given by
\begin{gather*}
p_n(x_1,\ldots,x_n) = \frac{1}{n!} \det_{i,j=1}^n K_n(x_i,x_j),
\end{gather*}
the mean counting probability is given by the density (note that $\tr K_n = n$)
\begin{gather*}
\rho_n(x) = n^{-1} K_n(x,x),
\end{gather*}
and the gap probabilities are given, by the inclusion-exclusion principle, in terms of a Fredholm determinant, namely
\begin{gather*}
E_n(J) = \prob(\{x_1,\ldots,x_n\} \cap J = \varnothing) = \det(I - \1_J K_n \1_J).
\end{gather*}
The various scaling limits are usually derived from an appropriate convergence of the kernel $K_n(x,y)$ by considering the large $n$ asymp\-totic of the eigenfunctions $\phi_j$, which can be technically quite involved\footnote{Based on the two-scale Plancherel--Rotach asymptotic of classical orthogonal polynomials or, methodologically more general, on the asymptotic of Riemann--Hilbert problems; see, e.g., Tracy and Widom~\cite{MR1257246,MR1266485}, Deift~\cite{MR1677884}, Lubinsky~\cite{MR2552113}, Johnstone~\cite{MR1863961, MR2485010}, Collins~\cite{MR2198015}, Forrester~\cite{MR2641363}, Anderson et al.~\cite{AGZ}, and Kuijlaars~\cite{uni}.}.

Borodin and Olshanski \cite{MR2326137} suggested, for discrete point processes, a dif\/ferent, conceptually and technically much simpler approach based on selfadjoint dif\/ference operators. We will show that their method, generalized to selfadjoint Sturm--Liouville operators, allows us to give a short and unif\/ied derivation of the various scaling limits for the random matrix ensembles with unitary invariance that are based on the classical orthogonal polynomials (Hermite, Laguerre, Jacobi).

\subsection*{The Borodin--Olshanski method}

The method proceeds along three steps: First, we identify the induced integral operator $K_n$ as the spectral projection (where we denote by $\chi_A$ the characteristic function of a Borel subset $A\subset \R$ and by $\chi_A(L_n)$ the application of that function to the selfadjoint operator $L_N$ in the sense of measurable functional calculus \cite[Theorem~VIII.6]{MR0493419})
\begin{gather*}
K_n = \1_{(-\infty,0)}(L_n)
\end{gather*}
of some selfadjoint ordinary dif\/ferential operator $L_n$ on $L^2(\Lambda)$. Any scaling of the point process by $x = \sigma_n \xi + \mu_n$ ($\sigma_n \neq 0$) yields, in turn, the induced rescaled operator
\begin{gather*}
\tilde K_n = \1_{(-\infty,0)}(\tilde L_n),
\end{gather*}
where $\tilde L_n$ is a selfadjoint dif\/ferential operator on $L^2(\tilde\Lambda_n)$, $\tilde \Lambda_n = (\tilde a_n,\tilde b_n)$.

Second, if $\tilde \Lambda_n \subset \tilde \Lambda = (\tilde a,\tilde b)$ with $\tilde a_n \to \tilde a$, $\tilde b_n \to \tilde b$, we aim for a selfadjoint operator $\tilde L$ on $L^2(\tilde \Lambda)$ with a core $C$ such that eventually $C \subset D(\tilde L_n)$ and
\begin{gather}\label{eq:Lpointwise}
\tilde L_n u \to \tilde Lu, \qquad u \in C.
\end{gather}
The point is that, if the test functions from $C$ are particularly nice, such a convergence is just a simple consequence of the \emph{locally uniform convergence of the coefficients} of the dif\/ferential operators~$\tilde L_n$~-- a~convergence that is, typically, an easy calculus exercise. Now, given (\ref{eq:Lpointwise}), the concept of \emph{ strong resolvent convergence} (see Theorem~\ref{thm:stolz}) immediately yields\footnote{By ``$\conv{s}$'' we denote the strong convergence of operators acting on~$L^2$.},
if $0 \not\in \sigma_{\rm pp}(\tilde L)$,
\begin{gather*}
\tilde K_n \1_{\tilde \Lambda_n} = \1_{(-\infty,0)}\big(\tilde L_n\big) \1_{\tilde \Lambda_n} \conv{s} \1_{(-\infty,0)}\big(\tilde L\big).
\end{gather*}

Third, we take an interval $J \subset \tilde\Lambda$, eventually satisfying $J \subset \tilde \Lambda_n$, such that the operator $\1_{(-\infty,0)}(\tilde L) \1_J$ is trace class with kernel $\tilde K(x,y)$ (which can be obtained from the generalized eigenfunction expansion of $\tilde L$, see Section~\ref{sect:genEFE}). Then, we immediately get the strong convergence
\begin{gather*}
\tilde K_n \1_J \conv{s} \tilde K \1_J.
\end{gather*}

\begin{Remark} Tao \cite[Section~3.3]{Tao} sketches the Borodin--Olshanski method, applied to the bulk and edge scaling of GUE, as a heuristic device. Because of the microlocal methods that he uses to calculate the projection $\1_{(-\infty,0)}(\tilde L)$, he puts his sketch under the headline ``The Dyson and Airy kernels of GUE via semiclassical analysis''.
\end{Remark}

\subsection*{Scaling limits and other modes of convergence} Given that one just has to establish the convergence of the coef\/f\/icients of a dif\/ferential operator (instead of an asymptotic of its eigenfunctions), the Borodin--Olshanski method is an extremely simple device to determine all the scalings $x=\sigma_n \xi + \mu_n$ that would yield some {\em meaningful} limit $\tilde K_n \1_J \to \tilde K \1_J$, namely in the strong operator topology. Other modes of convergence have been studied in the literature, ranging from some weak convergence of $k$-point correlation functions over convergence of the kernel functions to the convergence of gap probabilities, that is,
\begin{gather*}
\tilde E_n(J) = \det\big(I - \1_J \tilde K_n \1_J\big) \to \det\big(I - \1_J \tilde K \1_J\big) = \tilde E(J).
\end{gather*}
From a probabilistic point of view, the latter convergence is of particular interest and has been shown in at least three ways:
\begin{enumerate}
\item By Hadamard's inequality, convergence of the determinants follows directly from the locally uniform convergence of the kernels $K_n$ \cite[Lemma~3.4.5]{AGZ} and, for unbounded $J$, from additional large deviation estimates \cite[Lemma 3.3.2]{AGZ}. This way, the limit gap probabilities in the bulk and soft edge scaling limit of GUE can rigorously be established (see, e.g., Anderson et al.~\cite[Sections~3.5 and~3.7]{AGZ}). Johansson \cite[Lemma~3.1]{MR1737991} gives some general conditions on a scaling of the~$K_n$ such that the determinant converges to the soft edge of~GUE.
\item Since $A \mapsto \det(I-A)$ is continuous with respect to the trace class norm \cite[Theorem~3.4]{MR2154153}, $\tilde K_n \1_J \to \tilde K \1_J$ in trace class norm would generally suf\/f\/ice. Such a convergence can be proved by factorizing the trace class operators into Hilbert--Schmidt operators and obtaining the $L^2$-convergence of the factorized kernels once more from locally uniform convergence, see the work of Johnstone \cite{MR1863961, MR2485010} on the scaling limits of the LUE/Wishart ensembles and on the limits of the JUE/MANOVA ensembles.
\item Since $\1_J \tilde K_n \1_J$ and $\1_J \tilde K \1_J$ are selfadjoint and positive semi-def\/inite, yet another way is by observing that the convergence $\tilde K_n \1_J \to \tilde K \1_J$ in trace class norm is, for continuous kernels, equivalent \cite[Theorem~2.20]{MR2154153} to the combination of both, the convergence $\tilde K_n \1_J \to \tilde K \1_J$ in the {\em weak} operator topology and the convergence of the traces
\begin{gather}\label{eq:criterion}
\int_J \tilde K_n(\xi,\xi) d\xi \to \int_J \tilde K(\xi,\xi) d\xi.
\end{gather}
Once again, these convergences follow from locally uniform convergence of the kernels; see Deift \cite[Section~8.1]{MR1677884} for an application of this method to the bulk scaling limit of GUE.
\end{enumerate}

Since convergence in the strong operator topology implies convergence in the weak one, the Borodin--Olshanski method would thus establish the convergence of gap probabilities if we were only able to show condition (\ref{eq:criterion}) by some additional, similarly short and simply argument. Note that, by the ideal property of the trace class, condition (\ref{eq:criterion}) implies the same condition for all $J' \subset J$. We fall, however, short of conceiving a proof strategy for condition (\ref{eq:criterion}) that would be {\em independent} of all the laborious proofs of locally uniform convergence of the kernels.

\begin{Remark}
Contrary to the discrete case considered by Borodin and Olshanski, it is also not immediate to infer from the strong convergence of the induced integral operators the \emph{pointwise} convergence of the kernels. In Section~\ref{sect:density} we will need only a single such instance, namely
\begin{gather}\label{eq:T2}
\tilde K_n(0,0) \to \tilde K(0,0),
\end{gather}
to prove a limit law $\tilde \rho_n(t) dt \conv{w} \tilde \rho(t) dt$ for the mean counting probability. Using mollif\/ied Dirac deltas, pointwise convergence would generally follow, for continuously dif\/ferentiable $\tilde K_n(\xi,\eta)$, if we were able to bound, locally uniform, the gradient of $\tilde K_n(\xi,\eta)$. Then, by dominated convergence, criterion~(\ref{eq:criterion}) would already be satisf\/ied if we established an integrable bound of $\tilde K_n(\xi,\xi)$ on $J$. Since the scalings laws are, however, maneuvering just at the edge between trivial cases (i.e., zero limits) and divergent cases, it is conceivable that a proof of such bounds might not be signif\/icantly simpler than a proof of convergence of the gap probabilities itself.
\end{Remark}

\subsection*{The main result}

To prepare we recall how an integral kernel $K_n(x,y)$ is getting covariantly transformed in the presence of an af\/f\/ine coordinate change $x=\sigma_n \xi + \mu_n$, $y = \sigma_n \eta + \mu_n$: by invariance of the $1$-form
\begin{gather*}
K_n(x,y) dy = \tilde K_n(\xi,\eta) d\eta
\end{gather*}
the transformed kernel $\tilde K$ is given by
\begin{gather}\label{eq:affine}
\tilde K_n(\xi,\eta) = \sigma_n K_n(\sigma_n \xi + \mu_n,\sigma_n\eta+\mu_n).
\end{gather}

Using the Borodin--Olshanski method, we will prove the following general result for selfadjoint Sturm--Liouville operators; a result that adds a further class of problems to the \emph{universality} \cite{uni} of the Dyson, Airy, and Bessel kernel\footnote{For the def\/initions of the kernels $K_\text{\rm Dyson}$, $K_\text{\rm Airy}$, $K_\text{\rm Bessel}$ see \eqref{eq:dyson}, \eqref{eq:airy} and \eqref{eq:bessel}.} in the bulk, soft-edge, and hard-edge scaling limits.

\begin{Theorem}\label{thm:main} Let $\Lambda$ be one of the three domains $\Lambda = (-\infty,\infty)$, $\Lambda = (0,\infty)$, or $\Lambda = (0,1)$, and let $L_n$ be a selfadjoint realization on $L^2(\Lambda)$ of the formally selfadjoint Sturm--Liouville operator\/\footnote{Since, in this paper, we consider always a particular selfadjoint realization of a formal dif\/ferential operator, we will use the same letter to denote both.}
\begin{gather*}
 -\frac{d}{dx}\left(p(x)\frac{d}{dx}\right) + q_n(x) - \lambda_n
\end{gather*}
with coefficients $p, q_n\in C^\infty(\Lambda)$ such that $p(x) > 0$ for all $x\in\Lambda$. Assume that, for $t \in \Lambda$ and $n \to \infty$, there are asymptotic expansions
\begin{gather}\label{eq:kappascal}
n^{-2\kappa'}\lambda_n \sim \omega, \qquad
n^{-2\kappa'}q_n(n^\kappa t) \sim q_*(t), \qquad
n^{2\kappa''}p(n^\kappa t) \sim p_*(t) > 0,
\end{gather}
 with a remainder that is of order $O(n^{-1})$ locally uniform in $t$, and exponents normalized by
\begin{gather}\label{eq:K}
\kappa + \kappa' + \kappa'' = 1,\qquad \kappa\geq 0,
\end{gather}
where $\kappa< \frac23$ if $\Lambda = (0,1)$. Further assume that these expansions can be differentiated\/\footnote{We say that an expansion $f_n(t) - f(t) = O(1/n)$ can be dif\/ferentiated if $f_n'(t) - f'(t) = O(1/n)$.}, that the roots of $q_*(t)-\omega$ are simple, and that the spectral projection $K_n = \1_{(-\infty,0)}(L_n)$ is normalized by
\begin{gather*}
\tr K_n = n.
\end{gather*}
Let a scaling by $x = \sigma_n \xi + \mu_n$ induce the transformed projection kernel $\tilde K_n$ according to \eqref{eq:affine}.

Then, depending on particular choices of $\sigma_n$ and $\mu_n$, the following three scaling limits hold.
\begin{itemize}\itemsep=0pt
\item Bulk scaling limit: given $t \in \Lambda$ with $q_*(t) < \omega$, the scaling parameters
\begin{gather*}
\sigma_n = \frac{n^{\kappa-1}}{\tilde\rho(t)}, \qquad \mu_n = n^{\kappa} t,
 \end{gather*}
where
\begin{gather}\label{eq:rhotilde}
\tilde\rho(t) = \frac{1}{\pi}\sqrt{\frac{(\omega - q_*(t))_+}{p_*(t)}},
\end{gather}
yield, for a bounded interval $J$, the strong limit
\begin{gather*}
\tilde K_n \1_{J} \conv{s} K_\text{\rm Dyson} \1_{J}.
\end{gather*}
At $\xi=0$, the mean counting probability density $\rho_n(x) = n^{-1}K_n(x,x)$ transforms to the new
variable $t$ as
\begin{gather*}
\tilde \rho_n(t) = n^\kappa \rho(n^\kappa t).
\end{gather*}
 Under condition \eqref{eq:T2}, and if $\tilde \rho$ as defined in \eqref{eq:rhotilde} has unit mass on~$\Lambda$, there is the limit law
 \begin{gather*}
 \tilde\rho_n(t) dt \conv{w} \tilde\rho(t) dt.
 \end{gather*}
\item Soft-edge scaling limit: given $t_* \in \Lambda$ with $q_*(t_*) = \omega$, the scaling parameters
\begin{gather*}
\sigma_n = n^{\kappa-\frac23} \left(\frac{p_*(t_*)}{q'_*(t_*)}\right)^{1/3}, \qquad \mu_n = n^{\kappa} t_*,
\end{gather*}
yield, for $s\in\R$ and a $($not necessarily bounded$)$ interval $J\subset(s,\infty)$, the strong limit
\begin{gather*}
\tilde K_n \1_{J} \conv{s} K_\text{\rm Airy} \1_{J}.
\end{gather*}
\item Hard-edge scaling limit: given that $\Lambda=(0,\infty)$ or $\Lambda = (0,1)$ with
\begin{gather}\label{eq:hardassumpt}
p(0)=0, \qquad p'(0)>0,\qquad q_n(x) = q(x) = \gamma^2 x^{-1} + O(1), \qquad x\to 0,
\end{gather}
the scaling parameters
\begin{gather*}
\sigma_n = \frac{p'(0)}{4\omega n^{2\kappa'}}, \qquad \mu_n = 0,
\end{gather*}
yield, for a bounded interval $J\subset(0,\infty)$, the strong limit\/\footnote{Here, if $0 \leq \alpha < 1$, the selfadjoint realization $L_n$ is def\/ined by means of the boundary condition
\begin{gather}\label{eq:besselRB}
2x u'(x) - \alpha u(x) = o\big(x^{-\alpha/2}\big), \qquad x\to 0.
\end{gather}}
\begin{gather}\label{eq:alpha}
\tilde K_n \1_{J} \conv{s} \left. K_\text{\rm Bessel}^{(\alpha)} \1_{J}\right|_{\alpha = 2\gamma/\sqrt{p'(0)}}.
\end{gather}
\end{itemize}
\end{Theorem}

\begin{Remark}\label{rem:finiteTrace} Whether the interval $J$ in the strong operator limit $\tilde K_n \1_{J} \conv{s} K \1_{J}$ can be chosen unbounded or not depends on whether the limit operator $K \1_J$ is trace class or not (see the explicit formulae of the traces given in the appendix for each of the three limits): only in the former case we get a representation of the scaling limit in terms of a particular integral kernel, cf.\ Theorem~\ref{thm:kernel}. Note that it is impossible to use $J=\Lambda$ since $\tr K_n = n \to \infty$.
\end{Remark}

\subsection*{Outline of the paper}
The proof of Theorem~\ref{thm:main} is subject of Section~\ref{sect:method}. In Section~\ref{sect:OPS} we apply it to the classical orthogonal polynomials, which yields a short and unif\/ied derivation of the known formulae for the scaling limits for the classical random matrix ensembles with unitary invariance (GUE, LUE/Wishart, JUE/MANOVA). In fact, by a result of Tricomi, the only input needed is the weight function~$w$ of the orthogonal polynomials; from there one gets in a purely formula based fashion (by simple manipulations which can easily be coded in any computer algebra system), f\/irst, to the coef\/f\/icients~$p$ and~$q_n$ as well as to the eigenvalues~$\lambda_n$ of the Sturm--Liouville opera\-tor~$L_n$ and next, by applying Theorem~\ref{thm:main}, to the particular scaling limits.

To emphasize that our main result and its application is largely independent of concretely identifying the limit projection kernel $\tilde K$, we postpone this identif\/ication to Lemmas~\ref{lem:dyson}, \ref{lem:airy} and~\ref{lem:bessel}: there, using generalized eigenfunction expansions, we calculate the Dyson, Airy, and Bessel kernels directly from the limit dif\/ferential operator~$\tilde L$.

\section{Proof of the main result for Sturm--Liouville operators}\label{sect:method}

We start the proof of Theorem~\ref{thm:main} with some preparatory steps before we deal with the particular scaling limits. Since $L_n$ is a~selfadjoint realization on $L^2(\Lambda)$ of the Sturm--Liouville operator
\begin{gather*}
L_n = - \frac{d}{dx}\left(p(x) \frac{d}{dx} \right) + q_n(x) - \lambda_n
\end{gather*}
with $p, q_n\in C^\infty(\Lambda)$ and $p(x) > 0$ for $x \in \Lambda$, we have $C^\infty_0(\Lambda) \subset D(L_n)$.

\subsection*{Preparatory Step 1: transformation}

The scaling
\begin{gather*}
x = \sigma_n \xi + \mu_n, \qquad \sigma_n \neq 0,
\end{gather*}
maps $x\in\Lambda$ bijectively to $\xi \in \tilde\Lambda_n$. Since such an af\/f\/ine coordinate transform just induces a~{\em unitary equivalence} of integral and dif\/ferential operators, the spectral projection relation
\begin{gather*}
K_n = \chi_{(-\infty,0)}(L_n)
\end{gather*}
is left invariant if the kernel $K_n(x,y)$ is transformed according to~\eqref{eq:affine} and the dif\/ferential operator~$L_n$ is transformed using $d/dx = \sigma_n^{-1}d/d\xi$ as
\begin{gather*}
-\frac{1}{\sigma_n^2} \frac{d}{d\xi} \left(p(\sigma_n \xi + \mu_n) \frac{d}{d\xi}\right) +q_n(\sigma_n\xi+\mu_n) - \lambda_n.
\end{gather*}
Since the spectral projection to the negative part of the spectrum of a dif\/ferential operator is left invariant if we multiply that operator by some {\em positive} constant $\tau_n \sigma_n^2$, $\tau_n > 0$, we see that
\begin{gather*}
\tilde K_n = \chi_{(-\infty,0)}\big(\tilde L_n\big),
\end{gather*}
 where the transformed dif\/ferential operator is given f\/inally by
\begin{gather*}
\tilde L_n =-\frac{d}{d\xi} \left(\tilde p_n(\xi)\frac{d}{d\xi}\right) +\tilde q_n(\xi)
\end{gather*}
with coef\/f\/icients
\begin{gather}\label{eq:LtildeCoeff}
\tilde p_n(\xi) = \tau_n p(\sigma_n\xi + \mu_n),\qquad \tilde q_n(\xi) = \tau_n\sigma_n^2\left(q_n(\sigma_n\xi+\mu_n) - \lambda_n\right).
\end{gather}

\subsection*{Preparatory Step 2: strong operator limit}
Suppose the transformed domain $\tilde \Lambda_n = (a_n,b_n)$ satisf\/ies $a_n\to a$, $b_n\to b$. Then, with $\tilde\Lambda = (a,b)$ we have that, eventually, $C_0^\infty(\tilde\Lambda) \subset D(\tilde L_n)$. Further, suppose that the coef\/f\/icients of $\tilde L_n$ converge locally uniform in $\tilde \Lambda$ as (where the limit of $\tilde p_n(\xi)$ can be dif\/ferentiated)
\begin{gather*}
\tilde p_n(\xi) \to \tilde p(\xi),\qquad \tilde q_n(\xi) \to \tilde q(\xi),
\end{gather*}
such that the limit coef\/f\/icients $\tilde p>0$ and $\tilde q$ are smooth functions and
\begin{gather}\label{eq:Ltilde}
\tilde L = -\frac{d}{d\xi} \left( \tilde p(\xi) \frac{d}{d\xi}\right) + \tilde q(\xi)
\end{gather}
def\/ines a Sturm--Liouville operator that is essentially selfadjoint on $C_0^\infty(\tilde\Lambda) \subset L^2(\tilde\Lambda)$. Then, by dominated convergence, we get the convergence $\tilde L_n u \to \tilde L u$ in $L^2(\tilde\Lambda)$ for each test function $u$ in the core $C_0^\infty(\tilde\Lambda)$. Hence, by Theorem~\ref{thm:stolz} we have the strong operator convergence
\begin{gather*}
\tilde K_n \1_{J} \conv{s} \1_{(-\infty,0)}\big(\tilde L\big)\1_{J}
\end{gather*}
if $0 \not\in \sigma_{\rm pp}(L)$ and, eventually, $J \subset {\tilde \Lambda_n}$. In the particular cases considered in the following limit steps of the proof, the spectrum of $\tilde L$ is always absolutely continuous, that is, $\sigma_{\rm pp}(L)=\varnothing$. Finally, by Theorem~\ref{thm:kernel}, under the f\/inite trace condition mentioned already in Remark~\ref{rem:finiteTrace}, there is an integral kernel $\tilde K$ such that
\begin{gather*}
\1_{(-\infty,0)}\big(\tilde L\big)\1_{J} = \tilde K \1_J,
\end{gather*}
which f\/inishes the proof of a strong operator convergence in general.

\subsection*{Preparatory Step 3: Taylor expansions of the coef\/f\/icients}

\subsubsection*{The case $\boldsymbol{\mu_n = n^\kappa t}$}
Suppose that $t\in \Lambda$ is f\/ixed. The choice $\tau_n = 1/p(\mu_n)>0$ is then admissible and we get, if
\begin{gather*}
\sigma_n=o\big(n^{\kappa-1/2}\big),
\end{gather*}
from (\ref{eq:kappascal}), (\ref{eq:K}), and \eqref{eq:LtildeCoeff} by a Taylor expansion
\begin{gather}\label{eq:coeff1}
\tilde p_n(\xi) = 1 + o(1),\qquad \tilde q_n(\xi) = \frac{\sigma_n^2 n^{2-2\kappa}}{p_*(t)} \big(q_*(t) - \omega + \sigma_n n^{-\kappa} q'_*(t)\cdot \xi\big) + o(1),
\end{gather}
which holds locally uniform in $\xi \in \tilde\Lambda$ (where the expansion of $\tilde p_n(\xi)$ can be dif\/ferentiated).

\subsubsection*{The case $\boldsymbol{\mu_n = 0}$}

Suppose that the assumptions in \eqref{eq:hardassumpt} are met. If $\sigma_n\to 0^+$, the choice $\tau_n = 4\sigma_n/p'(0)>0$ is admissible and we get from~\eqref{eq:LtildeCoeff} by a Taylor expansion
\begin{gather}\label{eq:coeff2}
\tilde p_n(\xi) = 4\xi + o(1),\qquad
\tilde q_n(\xi) = \frac{4\gamma^2}{p'(0)\xi} - \frac{4\sigma_n\lambda_n}{p'(0)} + o(1),
\end{gather}
which holds locally uniform in $\xi \in \tilde\Lambda$ (where the expansion of $\tilde p_n(\xi)$ can be dif\/ferentiated).

\subsection*{Limit Step 1: bulk scaling limit}

If $q_*(t) \neq \omega$, by inserting
\begin{gather*}
\sigma_n = \sigma_n(t) = \pi n^{\kappa-1} \sqrt{\frac{p_*(t)}{|\omega - q_*(t)|}}
\end{gather*}
we read of\/f from \eqref{eq:coeff1} the limit coef\/f\/icients $\tilde p(\xi) = 1$ and $\tilde q(\xi) = - s \pi^2$, where $s = \sgn (\omega - q_*(t))$; that is,
the limit dif\/ferential operator \eqref{eq:Ltilde} is given by
\begin{gather*}
\tilde L = -\frac{d^2}{d\xi^2} - s \pi^2.
\end{gather*}
Note that, for the domains $\Lambda$ and the values of $\kappa$ considered, we have $\tilde\Lambda = (-\infty,\infty)$.

Lemma~\ref{lem:dyson} states that $\tilde L$ is essentially selfadjoint on $C^\infty_0(\tilde \Lambda)$ and that its unique selfadjoint extension has absolutely continuous spectrum: $\sigma(\tilde L) = \sigma_\text{ac}(\tilde L) = [-s \pi^2,\infty)$. Thus, for $s = -1$, the spectral projection $\1_{(-\infty,0)}(\tilde L)$ is zero. For $s=1$, the spectral projection can be calculated by a generalized eigenfunction expansion, yielding the \emph{Dyson} kernel~(\ref{eq:dyson}).

We will see in the next step that the dichotomy between $s=\pm 1$ is also ref\/lected in the structure of the support of the limit law $\tilde \rho$.

\subsection*{Limit Step~2: limit law}\label{sect:density}
The result for the bulk scaling limit allows, in passing, to calculate a limit law of the mean counting probability density $\rho_n(x) = n^{-1} K_n(x,x)$: we observe that $x = n^\kappa t$ transforms the density $\rho_n(x)$ into
\begin{gather*}
\tilde \rho_n(t) = n^{\kappa-1} K_n(n^\kappa t,n^\kappa t) = \frac{n^{\kappa-1}}{\sigma_n(t)} \tilde K_n(0,0)
= \frac{1}{\pi} \sqrt{\frac{|\omega - q_*(t)|}{p_*(t)}} \tilde K_n(0,0).
\end{gather*}
Thus, to get to a limit, we have to \emph{assume} condition (\ref{eq:T2}), so that a pointwise rendering of the bulk scaling limit just considered yields\footnote{The Iverson bracket $[S]$ stands for $1$ if the statement $S$ is true, $0$ otherwise.}
\begin{gather*}
\tilde K_n(0,0) \to [q_*(t) < \omega] K_\text{\rm Dyson}(0,0) = [q_*(t) < \omega].
\end{gather*}
This way we get
\begin{gather*}
\tilde \rho_n(t) \to \tilde\rho(t) = \frac{1}{\pi}\sqrt{\frac{(\omega - q_*(t))_+}{p_*(t)}}.
\end{gather*}
Hence, by Helly's selection theorem, the probability measure $\tilde \rho_n(t) dt$ converges vaguely to $\tilde\rho(t) dt$, which is, in general, just a sub-probability measure. If, however, it is checked that $\tilde\rho(t) dt$ has unit mass, the convergence is weak.

\subsection*{Limit Step 3: soft-edge scaling limit}\label{sect:soft}

If $q_*(t_*) = \omega$, by inserting\footnote{Note that, by the assumption made on the simplicity of the roots of $q_*(t)-\omega$, we have $q'_*(t_*) \neq 0$.}
\begin{gather*}
\sigma_n = \sigma_n(t_*) = n^{\kappa-2/3} \left(\frac{p_*(t_*)}{q'_*(t_*)}\right)^{1/3}
\end{gather*}
we read of\/f from \eqref{eq:coeff1} the limit coef\/f\/icients $\tilde p(\xi) = 1$ and $\tilde q(\xi) = \xi$;
that is, the limit dif\/ferential operator \eqref{eq:Ltilde} is
\begin{gather*}
\tilde L = -\frac{d^2}{d\xi^2} + \xi.
\end{gather*}
Note that, for the domains $\Lambda$ and the values of $\kappa$ considered, we have $\tilde\Lambda = (-\infty,\infty)$.

Lemma~\ref{lem:airy} states that $\tilde L$ is essentially selfadjoint on $C^\infty_0(\tilde \Lambda)$ and that its unique selfadjoint extension has absolutely continuous spectrum: $\sigma(\tilde L) = \sigma_\text{ac}(\tilde L)= (-\infty,\infty)$. The spectral projection can be calculated by a generalized eigenfunction expansion, yielding the \emph{Airy} kernel~(\ref{eq:airy}).

\subsection*{Limit Step 4: hard-edge scaling limit}\label{sect:hard}

For $\Lambda = (0,\infty)$ or $\Lambda=(0,1)$, we take a scaling
\begin{gather*}
x = \sigma_n \xi,
\end{gather*}
with $\sigma_n\to 0^+$ appropriately chosen, to explore the vicinity of the ``hard edge'' $x=0$; note that such a scaling yields $\tilde \Lambda = (0,\infty)$. We make the assumptions stated in (\ref{eq:hardassumpt}). By inserting
\begin{gather*}
\sigma_n = n^{-2\kappa'} \frac{p'(0)}{4\omega}
\end{gather*}
we read of\/f from \eqref{eq:coeff2}, using \eqref{eq:kappascal}, the limit coef\/f\/icients $\tilde p(\xi) = 4\xi$ and $\tilde q(\xi) = \alpha^2\xi^{-1}-1$, where~$\alpha$ is def\/ined as in \eqref{eq:alpha}; that is, the limit dif\/ferential operator \eqref{eq:Ltilde} is given by
\begin{gather*}
\tilde L = \left. -4\frac{d}{d\xi}\left(\xi \frac{d}{d\xi}\right) + \alpha^2 \xi^{-1} - 1\right|_{\alpha = 2\gamma/\sqrt{p'(0)}}.
\end{gather*}
If $\alpha \geq 1$, Lemma~\ref{lem:bessel} states that the limit $\tilde L$ is essentially selfadjoint on $C^\infty_0(\tilde \Lambda)$ and that the spectrum of its unique selfadjoint extension is absolutely continuous: $\sigma(\tilde L) = \sigma_\text{ac}(\tilde L) = [-1,\infty)$. The spectral projection can be calculated by a generalized eigenfunction expansion, yielding the \emph{Bessel} kernel~(\ref{eq:bessel}).

\begin{Remark}\label{rem:alpha1} The theorem also holds in the case $0 \leq \alpha < 1$ if the particular selfadjoint realization~$L_n$ is def\/ined by the boundary condition (\ref{eq:besselRB}), see Remark~\ref{rem:bessel}.
\end{Remark}

\section{Application to classical orthogonal polynomials}\label{sect:OPS}

In this section we apply Theorem~\ref{thm:main} to the kernels associated with the classical orthogonal polynomials, that is, the Hermite, Laguerre, and Jacobi polynomials. In random matrix theory, the thus induced determinantal processes are modeled by the spectra of the Gaussian unitary ensemble (GUE), the Wishart or Laguerre unitary ensemble (LUE), and the MANOVA (multivariate analysis of variance) or Jacobi unitary ensemble (JUE).

To prepare the study of the individual cases, we f\/irst discuss their common structure. Let~$P_n(x)$ be the sequence of classical orthogonal polynomials belonging to the weight function $w(x)$ on the (not necessarily bounded) interval~$(a,b)$. We normalize $P_n(x)$ such that $\langle \phi_n,\phi_n \rangle = 1$, where $\phi_n(x) = w(x)^{1/2} P_n(x)$. The functions $\phi_n$ form a complete orthogonal set in $L^2(a,b)$; conceptual proofs of the completeness can be found, e.g., in Andrews, Askey and Roy \cite{MR1688958} (Section~5.7 for the Jacobi polynomials, Section~6.5 for the Hermite and Laguerre polynomials).

By a result of Tricomi \cite[Section~10.7]{MR698780}, the $P_n(x)$ satisfy the eigenvalue problem
\begin{gather*}
-\frac{1}{w(x)} \frac{d}{dx}\left(p(x) w(x) \frac{d}{dx} P_n(x) \right) = \lambda_n P_n(x), \qquad \lambda_n = -n\big(r' + \tfrac{1}{2}(n+1)p''\big),
\end{gather*}
where $p(x)$ is a \emph{quadratic} polynomial\footnote{With the sign chosen such that $p(x)>0$ for $x \in (a,b)$.} and $r(x)$ a \emph{linear} polynomial such that
\begin{gather*}
\frac{w'(x)}{w(x)} = \frac{r(x)}{p(x)}.
\end{gather*}
In terms of $\phi_n$, a brief calculation shows that
\begin{gather*}
-\frac{d}{dx} \left(p(x) \frac{d}{dx} \phi_n(x) \right) + q(x) \phi_n(x) = \lambda_n \phi_n(x), \qquad q(x) = \frac{r(x)^2}{4p(x)} + \frac{r'(x)}{2}.
\end{gather*}
Therefore, by the completeness of the $\phi_n$, the formally selfadjoint Sturm--Liouville operator $L = -\frac{d}{dx} p(x) \frac{d}{dx} + q(x)$ has
a particular selfadjoint realization on $L^2(a,b)$ (which we continue to denote by the letter $L$) with spectrum
\begin{gather*}
\sigma(L) = \{\lambda_0,\lambda_1,\lambda_2,\ldots\}
\end{gather*}
and corresponding eigenfunctions~$\phi_n$. Hence, if the eigenvalues are, eventually, strictly increasing, the projection kernel (\ref{eq:prokernel}) def\/ines an integral operator $K_n$ with $\tr K_n =n $ such that, eventually,
\begin{gather*}
K_n = \1_{(-\infty,0)}(L_n),\qquad L_n = L-\lambda_n.
\end{gather*}
Note that this relation remains true if we choose to make some parameters of the weight $w$ (and, therefore, of the functions $\phi_j$) to depend on~$n$. For the scaling limits of $K_n$, we are now in the realm of Theorem~\ref{thm:main}: given the weight $w(x)$ as the only input all the other quantities can now be obtained simply by routine calculations.

\subsection*{Hermite polynomials}
The weight is $w(x) = e^{-x^2}$ on $\Lambda=(-\infty,\infty)$; hence
\begin{gather*}
p(x) = 1, \qquad r(x) = -2x, \qquad q(x) = x^2-1, \qquad \lambda_n = 2n,
\end{gather*}
and, therefore,
\begin{gather*}
\kappa = \kappa' = \tfrac{1}{2}, \qquad \kappa'' = 0, \qquad p_*(t) = 1, \qquad q_*(t) = t^2, \qquad\omega = 2.
\end{gather*}
Theorem~\ref{thm:main} is applicable and we directly read of\/f the following well-known scaling limits of the GUE (see, e.g., \cite[Chapter~3]{AGZ}):
\begin{itemize}\itemsep=0pt
\item bulk scaling limit: if $-\sqrt{2}< t <\sqrt{2}$, the transformation
\begin{gather*}
x = \frac{\pi \xi}{n^{1/2}\sqrt{2-t^2}} + n^{1/2} t
\end{gather*}
induces $\tilde K_n$ with a strong limit given by the \emph{Dyson kernel}; \item limit law: the transformation $x = n^{1/2} t$ induces the mean counting probability density $\tilde \rho_n$ with a weak limit given by the \emph{Wigner semicircle law}
\begin{gather*}
\tilde\rho(t) = \frac{1}{\pi} \sqrt{(2-t^2)_+};
\end{gather*}
\item soft-edge scaling limit: the transformation
\begin{gather*}
x = \pm\big(2^{-1/2}n^{-1/6}\xi + \sqrt{2n}\big)
\end{gather*}
induces $\tilde K_n$ with a strong limit given by the \emph{Airy kernel}.
\end{itemize}

\subsection*{Laguerre polynomials}
The weight is $w(x) = x^\alpha e^{-x}$ on $\Lambda=(0,\infty)$; hence
\begin{gather*}
p(x) = x, \qquad r(x) = \alpha -x, \qquad q(x) = \frac{(\alpha-x)^2}{4x} - \frac{1}{2}, \qquad \lambda_n = n.
\end{gather*}
In random matrix theory, the corresponding determinantal point process is modeled by the spectra of complex $n \times n$ Wishart matrices with a dimension parameter $m \geq n$; the Laguerre parameter $\alpha$ is then given by $\alpha = m - n \geq 0$. Of particular interest in statistics \cite{MR1863961} is the simultaneous limit $m,n \to \infty$ with
\begin{gather*}
\frac{m}{n} \to \theta \geq 1,
\end{gather*}
for which we get
\begin{gather*}
\kappa = 1, \qquad \kappa' = \frac{1}{2}, \qquad \kappa'' = -\frac{1}{2}, \qquad p_*(t) = t, \qquad q_*(t) = \frac{(\theta-1-t)^2}{4t}, \qquad\omega = 1.
\end{gather*}
Note that
\begin{gather*}
\omega - q_*(t) = \frac{(t_+ - t)(t-t_-)}{4t},\qquad t_\pm = \big(\sqrt{\theta} \pm 1\big)^2.
\end{gather*}
Theorem~\ref{thm:main} is applicable and we directly read of\/f the following well-known scaling limits of the Wishart ensemble~\cite{MR1863961}:
\begin{itemize}\itemsep=0pt
\item bulk scaling limit: if $t_- < t <t_+$,
\begin{gather*}
x = \frac{2\pi t \xi}{\sqrt{(t_+-t)(t-t_-)}} + n t
\end{gather*}
induces $\tilde K_n$ with a strong limit given by the \emph{Dyson kernel};
\item limit law: the scaling $x = n t$ induces the mean counting probability density $\tilde \rho_n$ with a~weak limit given by the \emph{Marchenko--Pastur law}
\begin{gather*}
\tilde\rho(t) = \frac{1}{2\pi t} \sqrt{((t_+-t)(t-t_-))_+};
\end{gather*}
\item soft-edge scaling limit: with signs chosen consistently as either $+$ or $-$,
\begin{gather}\label{eq:johnstone}
x = \pm n^{1/3} \theta^{-1/6}t_{\pm}^{2/3}\xi + n t_{\pm}
\end{gather}
induces $\tilde K_n$ with a strong limit given by the \emph{Airy kernel}.
\end{itemize}

\begin{Remark}
The scaling (\ref{eq:johnstone}) is better known in the asymptotically equivalent form
\begin{gather*}
x = \sigma \xi + \mu,\qquad \mu = \big(\sqrt{m}\pm\sqrt{n}\big)^2,\qquad \sigma = \big(\sqrt{m}\pm\sqrt{n}\big)\left(\frac{1}{\sqrt{m}}\pm \frac{1}{\sqrt{n}}\right)^{1/3},
\end{gather*}
which is obtained from (\ref{eq:johnstone}) by replacing $\theta$ with $m/n$, see \cite[p.~305]{MR1863961}.
\end{Remark}

In the case $\theta = 1$, which implies $t_- = 0$, the lower soft-edge scaling (\ref{eq:johnstone}) breaks down and has to be replaced by a scaling at the \emph{hard} edge:
\begin{itemize}\itemsep=0pt
\item hard-edge scaling limit: if $\alpha = m - n$ is a constant\footnote{By\label{foot:alpha1} Remark~\ref{rem:alpha1}, there is no need to restrict ourselves to $\alpha\geq 1$: since $\phi_n(x) = x^\alpha \tilde\phi_n(x)$ with $\tilde\phi_n(x)$ extending smoothly to $x=0$, we have, for $\alpha\geq 0$,
\begin{gather*}
x^{\alpha/2} (2x \phi_n'(x) -\alpha \phi_n(x)) = 2x^{1+\alpha} \tilde\phi_n'(x) = O(x),\qquad x\to 0.
\end{gather*}
Hence, the selfadjoint realization $L_n$ is compatible with the boundary condition (\ref{eq:besselRB}).}, $x = \xi/(4n)$ induces $\tilde K_n$ with a strong limit given by the \emph{Bessel kernel} $K_\text{\rm Bessel}^{(\alpha)}$.
\end{itemize}

\subsection*{Jacobi polynomials}
The weight is $w(x) = x^\alpha (1-x)^\beta$ on $\Lambda=(0,1)$; hence
\begin{gather*}
p(x) = x(1-x), \qquad r(x) = \alpha - (\alpha+\beta) x, \qquad q(x) = \frac{(\alpha - (\alpha+\beta) x)^2}{4x(1-x)} - \frac{\alpha+\beta}{2},
\end{gather*}
and
\begin{gather*}
\lambda_n = n(n+\alpha+\beta+1).
\end{gather*}
In random matrix theory, the corresponding determinantal point process is modeled by the spectra of complex $n \times n$ MANOVA matrices with dimension parameters $m_1, m_2 \geq n$; the Jacobi parameters $\alpha$, $\beta$ are then given by $\alpha = m_1 - n \geq 0$ and $\beta = m_2 -n \geq 0$. Of particular interest in statistics \cite{MR2485010} is the simultaneous limit $m_1,m_2,n \to \infty$ with
\begin{gather*}
\frac{m_1}{m_1+m_2} \to \theta \in (0,1),\qquad \frac{n}{m_1+m_2} \to \tau \in (0,1/2],
\end{gather*}
for which we get
\begin{gather*}
\kappa = \kappa'' = 0, \qquad\!\!\! \kappa' = 1, \qquad\!\! p_*(t) = t(1-t), \qquad\!\!\! q_*(t) = \frac{(\theta-\tau-(1-2\tau)t)^2}{4\tau^2 t(1-t)}, \qquad\!\!\! \omega = \frac{1-\tau}{\tau}.
\end{gather*}
Note that
\begin{gather*}
\omega - q_*(t) = \frac{(t_+ - t)(t-t_-)}{4\tau^2 t(1-t)},\qquad t_\pm = \left(\sqrt{\theta(1-\tau)}\pm\sqrt{\tau(1-\theta)}\right)^2.
\end{gather*}
Theorem~\ref{thm:main} is applicable and we directly read of\/f the following (less well-known) scaling limits of the MANOVA ensemble \cite{MR2198015,MR2485010}:
\begin{itemize}\itemsep=0pt
\item bulk scaling limit: if $t_- < t <t_+$,
\begin{gather*}
x = \frac{2\pi \tau t(1-t) \xi}{n\sqrt{(t_+-t)(t-t_-)}} + t
\end{gather*}
induces $\tilde K_n$ with a strong limit given by the \emph{Dyson kernel}; \item limit law: (because of $\kappa=0$ there is no transformation here) the mean counting probability density $\rho_n$ has a weak limit given by the law~\cite{MR585695}
\begin{gather*}
\rho(t) = \frac{1}{2\pi \tau t(1-t)} \sqrt{((t_+-t)(t-t_-))_+};
\end{gather*}
\item soft-edge scaling limit: with signs chosen consistently as either $+$ or $-$,
\begin{gather}\label{eq:johnstone2}
x = \pm n^{-2/3} \frac{(\tau t_{\pm}(1-t_\pm))^{2/3}}{(\tau \theta(1-\tau)(1-\theta))^{1/6}} \xi + t_{\pm}
\end{gather}
induces $\tilde K_n$ with a strong limit given by the \emph{Airy kernel}.
\end{itemize}
\begin{Remark} Johnstone \cite[p.~2651]{MR2485010} gives the soft-edge scaling in terms of a trigonometric parametrization of $\theta$ and $\tau$. By putting
\begin{gather*}
\theta = \sin^2\frac{\phi}{2},\qquad \tau = \sin^2\frac{\psi}{2},
\end{gather*}
we immediately get
\begin{gather*}
t_\pm = \sin^2\frac{\phi\pm\psi}{2}
\end{gather*}
and (\ref{eq:johnstone2}) becomes
\begin{gather*}
x = \pm \sigma_{\pm} \xi + t_\pm, \qquad \sigma_\pm = n^{-2/3}\left(\frac{\tau^2 \sin^4(\phi\pm\psi)}{4\sin\phi\sin\psi}\right)^{1/3}.
\end{gather*}
\end{Remark}

In the case $\theta =\tau = 1/2$, which is equivalent to $m_1/n, m_2/n \to 1$, we have $t_- = 0$ and $t_+=1$. Hence, the lower and the upper soft-edge scaling (\ref{eq:johnstone2}) break down and have to be replaced by a scaling at the \emph{hard} edges:
\begin{itemize}\itemsep=0pt
\item hard-edge scaling limit: if $\alpha = m_1 - n$, $\beta = m_2 -n$ are constants\footnote{For the cases $0\leq \alpha < 1$ and $0 \leq \beta <1$, see the justif\/ication of the limit given in footnote~\ref{foot:alpha1}.}, $x = \xi/(4n^2)$ induces~$\tilde K_n$ with a strong limit given by the \emph{Bessel kernel} $K_\text{\rm Bessel}^{(\alpha)}$; by symmetry, the \emph{Bessel kernel} $K_\text{\rm Bessel}^{(\beta)}$ is obtained for $x = 1 - \xi/(4n^2)$.
\end{itemize}

\renewcommand{\thesection}{A}
\section{Appendices}

\subsection{Generalized strong convergence}\label{sect:StrongResolvent}

The notion of {\em strong resolvent convergence} \cite[Section~9.3]{MR566954} links the convergence of dif\/ferential operators, tested for an appropriate class of smooth functions, to the strong convergence of their spectral projections. We recall a slight generalization of that concept, which allows the underlying Hilbert space to vary.

Specif\/ically we consider, on an interval $(a,b)$ (not necessarily bounded) and on a sequence of subintervals $(a_n,b_n) \subset (a,b)$ with $a_n \to a$ and $b_n\to b$, selfadjoint operators
\begin{gather*}
L\colon \ D(L) \subset L^2(a,b) \to L^2(a,b), \qquad L_n \colon \ D(L_n) \subset L^2(a_n,b_n) \to L^2(a_n,b_n).
\end{gather*}
By means of the natural embedding (that is, extension by zero) we take $L^2(a_n,b_n) \subset L^2(a,b)$; the multiplication operator induced by the characteristic function $\1_{(a_n,b_n)}$, which we will denote by the same symbol, constitutes the orthogonal projection of $L^2(a,b)$ onto $L^2(a_n,b_n)$. Following Stolz and Weidmann \cite[Section~2]{MR1244968}, we say that $L_n$ converges to $L$ in the sense of \emph{generalized strong convergence} (gsc), if for some $z \in \C\setminus\R$, and hence, a forteriori, for all such~$z$,
\begin{gather*}
R_z(L_n) \1_{(a_n,b_n)} \conv{s} R_z(L),\qquad n\to\infty,
\end{gather*}
in the strong operator topology of $L^2(a,b)$.\footnote{We denote by $R_z(L) = (L- z)^{-1}$ the resolvent of an operator $L$.}

\begin{Theorem}[Stolz and Weidmann \protect{\cite[Theorem~4/5]{MR1244968}}]\label{thm:stolz} Let the selfadjoint operators~$L_n$ and~$L$ satisfy the assumptions stated above and let $C$ be a core of $L$ such that, eventually, $C \subset D(L_n)$.
\begin{itemize}\itemsep=0pt
\item[$(i)$] If $L_n u \to Lu$ for all $u \in C$, then $L_n \conv{gsc} L$.
\item[$(ii)$] If $L_n \conv{gsc} L$ and if the endpoints of the interval $\Delta \subset \R$ do not belong to the pure point spectrum $\sigma_{\text{\rm pp}}(L)$ of $L$, the spectral projections to $\Delta$ converge as
\begin{gather*}
\1_\Delta(L_n) \1_{(a_n,b_n)} \conv{s} \1_\Delta(L).
\end{gather*}
\end{itemize}
\end{Theorem}

\subsection{Generalized eigenfunction expansion of Sturm--Liouville operators}\label{sect:genEFE}

Let $L$ be a formally selfadjoint Sturm--Liouville operator on the interval $(a,b)$,
\begin{gather*}
L u = -(p u')' + q u,
\end{gather*}
with smooth coef\/f\/icient functions $p>0$ and $q$. We have the \emph{limit point case} (LP) at the boundary point $a$ if there is some $c \in (a,b)$ and some $z \in \C$ such that there exists at least one solution of $(L-z)u=0$ in $(a,b)$ for which $u \not\in L^2(a,c)$; otherwise, we have the {\em limit circle case} (LC) at~$a$. According to the Weyl alternative \cite[Theorem~8.27]{MR566954}, in the LP case there exists actually for {\em all} $c\in (a,b)$ and {\em all} $z\in \C$ at least one solution of $(L-z)u=0$ in $(a,b)$ for which $u\not\in L^2(a,c)$; yet, if $z \in \C\setminus\R$, there is a one-dimensional space of solutions $u$ of $(L-z)u=0$ for which there is nevertheless $u \in L^2(a,c)$. The same structure and notion applies to the boundary point~$b$.

\begin{Theorem}\label{thm:weyl} Let $L$ be a formally selfadjoint Sturm--Liouville operator on the interval $(a,b)$ as defined above. If there is the LP case at $a$ and $b$, then $L$ is essentially self-adjoint on the domain $C^\infty_0(a,b)$ and, for $z \in \C\setminus\R$, the resolvent $R_z(L) = (L-z)^{-1}$ of its unique selfadjoint extension (which we continue to denote by the letter $L$) is of the form
\begin{gather}\label{eq:green1}
R_z(L)\phi(x) = \frac{1}{W(u_a,u_b)} \left(u_b(x)\int_a^x u_a(y)\phi(y) dy + u_a(x) \int_x^b u_b(y) \phi(y) dy\right).
\end{gather}
Here $u_a$ and $u_b$ are the non-vanishing solutions of the equation $(L-z)u=0$, uniquely determined up to a factor by the conditions $u_a \in L^2(a,c)$ and $u_b \in L^2(c,b)$ for some $c \in (a,b)$, and~$W$ denotes the Wronskian
\begin{gather*}
W(u_a,u_b) = p(x) (u_a'(x)u_b(x) - u_a(x) u_b'(x)),
\end{gather*}
which is a constant for $x \in (a,b)$.
\end{Theorem}

A more general formulation of this theorem, which includes also the LC case, can be found, e.g., in \cite[Theorem~8.26/8.29]{MR566954}; see \cite[pp.~41--42]{MR923320} for a proof that $C^\infty_0(a,b)$ is a core of $L$ if the coef\/f\/icients are smooth. In the following, we write (\ref{eq:green1}) brief\/ly in the form
\begin{gather*}
R_z(L) \phi(x) = \int_a^b G_z(x,y)\phi(y) dy
\end{gather*}
with the {\em Green's kernel}
\begin{gather*}
G_z(x,y) = \frac{1}{W(u_a,u_b)} \begin{cases}
u_b(x)u_a(y), & x > y, \\
u_a(x)u_b(y), & \text{otherwise}.
\end{cases}
\end{gather*}
If the imaginary part of $G_z(x,y)$ has f\/inite boundary values as $z$ approaches the real line from above, there is a simple formula for the spectral projection associated with $L$ that often applies if the spectrum of $L$ is absolutely continuous.

\begin{Theorem}\label{thm:kernel}\quad
\begin{enumerate}\itemsep=0pt
\item[$(i)$] Assume that there exits, as $\epsilon \to 0^+$, the limit
\begin{gather*}
\pi^{-1} \Im G_{\lambda+i\epsilon}(x,y) \to K_\lambda(x,y),
\end{gather*}
locally uniform in $x,y \in (a,b)$ for each $\lambda \in \R$ except for some isolated points $\lambda$ for which the limit
is replaced by
\begin{gather*}
\epsilon \Im G_{\lambda+i\epsilon}(x,y) \to 0.
\end{gather*}
Then the spectrum is absolutely continuous, $\sigma(L) = \sigma_{\text{ac}}(L)$, and, for a Borel set $\Delta$,
\begin{gather}\label{eq:projkern}
\langle \1_\Delta(L) \phi,\psi\rangle = \int_\Delta \langle K_\lambda \phi,\psi\rangle d\lambda,\qquad \phi,\psi \in C^\infty_0(a,b).
\end{gather}
\item[$(ii)$] Assume further, for some $(a',b')\subset(a,b)$, that
\begin{gather*}
\int_{a'}^{b'} \int_{a'}^{b'} \left(\int_\Delta |K_\lambda(x,y)| d\lambda\right)^2 dx dy < \infty.
\end{gather*}
Then $\1_\Delta(L)\1_{(a',b')}$ is a Hilbert--Schmidt operator on $L^2(a,b)$ with kernel
\begin{gather*}
 \1_{(a',b')}(y) \int_\Delta K_\lambda(x,y) d\lambda.
\end{gather*}
If $\int_\Delta K_\lambda(x,y) d\lambda$ is a continuous function of $x,y \in (a',b')$, $\1_\Delta(L)\1_{(a',b')}$ is a trace class operator with trace
\begin{gather*}
\tr \1_\Delta(L)\1_{(a',b')} = \int_{a'}^{b'} \int_\Delta K_\lambda(x,x) d\lambda dx.
\end{gather*}
\end{enumerate}
\end{Theorem}

\begin{proof} With $E$ denoting the spectral resolution of the selfadjoint operator~$L$, we observe that, for a given $\phi \in C^\infty_0(a,b)$, the Borel--Stieltjes transform of the positive measure $\mu_\phi(\lambda)=\langle E(\lambda) \phi,\phi\rangle$ can be simply expressed in terms of the resolvent as follows, see \cite[Section~32.1]{MR1892228}:
\begin{gather*}
\int_{-\infty}^\infty \frac{d\mu_\phi(\lambda)}{\lambda-z} = \langle R_z(L) \phi,\phi\rangle.
\end{gather*}
If we take $z = \lambda + i \epsilon$ and let $\epsilon \to 0^+$, we obtain by the locally uniform convergence of the integral kernel of~$R_z$ that there exits either the limit
\begin{gather*}
\pi^{-1} \Im \langle R_{\lambda+i\epsilon} (L) \phi,\phi\rangle \to \langle K_\lambda \phi,\phi\rangle
\end{gather*}
or, at isolated points $\lambda$,
\begin{gather*}
\epsilon \Im \langle R_{\lambda+i\epsilon} (L) \phi,\phi\rangle \to 0.
\end{gather*}
By a theorem of de la Vall\'ee--Poussin \cite[Theorem~11.6(ii/iii)]{MR2154153}, the singular part of $\mu_\phi$ va\-nishes, $\mu_{\phi,\text{sing}} = 0$; by Plemelj's reconstruction the absolutely continuous part satisf\/ies \cite[Theo\-rem~11.6(iv)]{MR2154153}
\begin{gather*}
d\mu_{\phi,\text{ac}}(\lambda) = \langle K_\lambda \phi,\phi\rangle d\lambda.
\end{gather*}
Since $C^\infty_0(a,b)$ is dense in $L^2(a,b)$, approximation shows that $E_\text{sing}=0$, that is, $\sigma(L)=\sigma_\text{ac}(L)$. Since $\langle \1_\Delta(L) \phi,\phi\rangle = \int_\Delta d\mu_\phi(\lambda)$, we thus get, by the symmetry of the bilinear expressions, the representation~(\ref{eq:projkern}), which f\/inishes the proof of~(i). The Hilbert--Schmidt part of part (ii) follows using the Cauchy--Schwarz inequality and Fubini's theorem and yet another density argument; the trace class part follows from \cite[Theo\-rem~IV.8.3]{MR1744872} since $\1_{(a',b')}\1_\Delta(L)\1_{(a',b')}$ is a~selfadjoint, positive-semidef\/inite operator.
\end{proof}

We apply this theorem to the spectral projections used in the proof of Theorem~\ref{thm:main}. The f\/irst two examples could have been dealt with by Fourier techniques \cite[Section~3.3]{Tao}; applying, however, the same method in all the examples renders the approach more systematic.

\begin{Example}[Dyson kernel]\label{exam:dyson} Consider $L u = -u''$ on $(-\infty,\infty)$. Since $u\equiv 1$ is a solution of $L u = 0$, both endpoints are LP; for a given $\Im z >0$ the solutions $u_a$ ($u_b$) of $(L -z) u= 0$ being~$L^2$ at~$-\infty$~($\infty$) are spanned by
\begin{gather*}
u_a(x) = e^{-i x \sqrt{z}},\qquad u_b(x) = e^{i x \sqrt{z}}.
\end{gather*}
Thus, Theorem~\ref{thm:weyl} applies: $L$ is essentially selfadjoint on $C^\infty_0(-\infty,\infty)$, the resolvent of its unique selfadjoint extension is represented, for $\Im z>0$, by the Green's kernel
\begin{gather*}
G_z(x,y) = \frac{i}{2\sqrt{z}} \begin{cases}
e^{i(x-y)\sqrt{z}}, & x > y, \\
e^{-i(x-y)\sqrt{z}}, & \text{otherwise}.
\end{cases}
\end{gather*}
For $\lambda > 0$ there is the limit
\begin{gather*}
\pi^{-1} \Im G_{\lambda+i0}(x,y) = K_\lambda(x,y) = \frac{\cos\big((x-y)\sqrt{\lambda}\big)}{2\pi\sqrt{\lambda}},
\end{gather*}
for $\lambda<0$ the limit is zero; both limits are locally uniform in $x,y \in \R$. For $\lambda = 0$ there would be divergence, but we obviously have
\begin{gather*}
 \epsilon \Im G_{i\epsilon}(x,y) \to 0, \qquad \epsilon\to 0^+,
\end{gather*}
locally uniform in $x,y \in \R$. Hence, Theorem~\ref{thm:kernel} applies: $\sigma(L)=\sigma_\text{ac}(L)=[0,\infty)$ and~(\ref{eq:projkern}) holds
for each Borel set $\Delta \subset \R$. Given a bounded interval $(a,b)$, we may estimate for the specif\/ic choice $\Delta = (-\infty,\pi^2)$ that
\begin{gather*}
\int_{a}^{b} \int_{a}^{b} \left(\int_{-\infty}^{\pi^2} |K_\lambda(x,y)| d\lambda\right)^2 dx dy\\
\qquad{}= \int_{a}^{b} \int_{a}^{b} \left(\int_0^{\pi^2} \left|\frac{\cos\big((x-y)\sqrt{\lambda}\big)}{2\pi\sqrt{\lambda}}\right| d\lambda\right)^2 dx dy
 \leq \left(\int_a^b \int_0^{\pi^2}\frac{d\lambda}{2\pi\sqrt{\lambda}}\right)^2= (b-a)^2.
\end{gather*}
Therefore, Theorem~\ref{thm:kernel} yields that $\1_{(-\infty,\pi^2)}(L)\1_{(a,b)}$ is Hilbert--Schmidt with the \emph{Dyson kernel}
\begin{gather*}
\int_{-\infty}^{\pi^2} K_\lambda(x,y) d\lambda = \int_{0}^{\pi^2} \frac{\cos\big((x-y)\sqrt{\lambda}\big)}{2\pi\sqrt{\lambda}} d\lambda = \frac{\sin(\pi(x-y))}{\pi(x-y)},
\end{gather*}
restricted to $x,y \in (a,b)$. Here, the last equality is simply obtained from
\begin{gather*}
(x-y) \int_{0}^{\pi^2} \frac{\cos\big((x-y)\sqrt{\lambda}\big)}{2\sqrt{\lambda}} d\lambda = \int_{0}^{\pi^2} \frac{d}{d\lambda} \sin\big((x-y)\sqrt{\lambda}\big) d\lambda = \sin(\pi(x-y)).
\end{gather*}
Since the resulting kernel is continuous for $x,y \in (a,b)$, Theorem~\ref{thm:kernel} gives that $\1_{(-\infty,\pi^2)}(L)\1_{(a,b)}$ is a trace class operator with trace
\begin{gather*}
\tr \1_{(-\infty,\pi^2)}(L)\1_{(a,b)} = b-a.
\end{gather*}

To summarize, we have thus obtained the following lemma.
\begin{Lemma}\label{lem:dyson} The operator $L u = -u''$ is essentially selfadjoint on $C^\infty_0(-\infty,\infty)$. The spectrum of its unique selfadjoint
extension is
\begin{gather*}
\sigma(L)=\sigma_{\rm ac}(L)=[0,\infty).
\end{gather*}
Given $(a,b)$ bounded, $\1_{(-\infty,\pi^2)}(L) \1_{(a,b)}$ is trace class with trace $b-a$ and kernel
\begin{gather}\label{eq:dyson}
K_{\text{\rm Dyson}}(x,y) = \int_{0}^{\pi^2} \frac{\cos\big((x-y)\sqrt{\lambda}\big)}{2\pi\sqrt{\lambda}} d\lambda = \frac{\sin(\pi(x-y))}{\pi(x-y)}.
\end{gather}
\end{Lemma}
\end{Example}

\begin{Example}[Airy kernel]\label{exam:airy} Consider the dif\/ferential operator $L u = -u'' + x u$ on $(-\infty,\infty)$. Since the specif\/ic solution $u(x) = \Bi(x)$ of $L u = 0$ is not locally $L^2$ at each of the endpoints, both endpoints are LP. For a given $\Im z >0$ the solutions $u_a$ ($u_b$) of $(L -z) u= 0$ being~$L^2$ at~$-\infty$~($\infty$) are spanned by \cite[equation~(10.4.59-64)]{MR0167642}
\begin{gather*}
u_a(x) = \Ai(x-z) - i \Bi(x-z),\qquad u_b(x) = \Ai(x-z).
\end{gather*}
Thus, Theorem~\ref{thm:weyl} applies: $L$ is essentially selfadjoint on $C^\infty_0(-\infty,\infty)$, the resolvent of its unique selfadjoint extension is represented, for $\Im z>0$, by the Green's kernel
\begin{gather*}
G_z(x,y) = i \pi \begin{cases}
\Ai(x-z)\left(\Ai(y-z)-i \Bi(y-z)\right), & x > y, \\
\Ai(y-z)\left(\Ai(x-z)-i \Bi(x-z)\right), & \text{otherwise}.
\end{cases}
\end{gather*}
For $\lambda \in\R$ there is thus the limit
\begin{gather*}
\pi^{-1} \Im G_{\lambda+i0}(x,y) = K_\lambda(x,y) = \Ai(x-\lambda)\Ai(y-\lambda),
\end{gather*}
locally uniform in $x,y \in \R$. Hence, Theorem~\ref{thm:kernel} applies: $\sigma(L)=\sigma_\text{ac}(L)=\R$ and (\ref{eq:projkern}) holds for each Borel set $\Delta \subset \R$. Given $s>-\infty$, we may estimate for the specif\/ic choice $\Delta = (-\infty,0)$ that
\begin{gather*}
\left(\int_{s}^{\infty} \int_{s}^{\infty} \left(\int_\Delta |K_\lambda(x,y)| d\lambda\right)^2 dx dy\right)^{1/2}
\leq \int_s^\infty \int_{0}^{\infty}\Ai(x+\lambda)^2 d\lambda dx = \tau(s)
\end{gather*}
with
\begin{gather*}
\tau(s) = \frac{1}{3} \left(2 s^2 \Ai (s)^2-2 s \Ai '(s)^2-\Ai (s)
 \Ai '(s)\right).
\end{gather*}
Therefore, Theorem~\ref{thm:kernel} yields that $\1_{(-\infty,0)}(L)\1_{(s,\infty)}$ is Hilbert--Schmidt with the \emph{Airy kernel}
\begin{gather*}
\int_{-\infty}^{0} K_\lambda(x,y) d\lambda = \int_0^\infty \Ai(x+\lambda)\Ai(y+\lambda) d\lambda=\frac{\Ai(x)\Ai'(y)-\Ai'(x)\Ai(y)}{x-y},
\end{gather*}
restricted to $x,y \in (s,\infty)$. Here, the last equality is obtained from a Christof\/fel--Darboux type of argument: First, we use the underlying dif\/ferential equation,
\begin{gather*}
x \Ai(x+\lambda) = \Ai''(x+\lambda) - \lambda \Ai(x+\lambda),
\end{gather*}
and partial integration to obtain
\begin{gather*}
x \int_{0}^\infty\! \Ai(x+\lambda)\Ai(y+\lambda) d\lambda
= \int_{0}^\infty\! \Ai''(x+\lambda)\Ai(y+\lambda) d\lambda - \int_{0}^\infty\! \lambda \Ai(x+\lambda)\Ai(y+\lambda) d\lambda\\
\qquad{}= - \Ai'(x)\Ai(y) - \int_{0}^\infty \Ai'(x+\lambda)\Ai'(y+\lambda) d\lambda - \int_{0}^\infty \lambda \Ai(x+\lambda)\Ai(y+\lambda) d\lambda.
\end{gather*}
Next, we exchange the roles of $x$ and $y$ and substract to get the assertion. Since the resulting kernel is continuous, Theorem~\ref{thm:kernel} gives that $\1_{(-\infty,0)}(L)\1_{(s,\infty)}$ is a trace class operator with trace
\begin{gather*}
\tr \1_{(-\infty,0)}(L)\1_{(s,\infty)} = \tau(s) \to \infty, \qquad s\to -\infty.
\end{gather*}
\end{Example}

To summarize, we have thus obtained the following lemma.
\begin{Lemma}\label{lem:airy} The differential operator $L u = -u'' + x u$ is essentially selfadjoint on $C^\infty_0(-\infty,\infty)$. The spectrum of its unique selfadjoint extension is
\begin{gather*}
\sigma(L)=\sigma_{\rm ac}(L)=(-\infty,\infty).
\end{gather*}
Given $s>-\infty$, the operator $\1_{(-\infty,0)}(L) \1_{(s,\infty)}$ is trace class with kernel
\begin{gather}\label{eq:airy}
K_\text{\rm Airy}(x,y) = \int_0^\infty \Ai(x+\lambda)\Ai(y+\lambda) d\lambda = \frac{\Ai(x)\Ai'(y)-\Ai'(x)\Ai(y)}{x-y}.
\end{gather}
\end{Lemma}

\begin{Example}[Bessel kernel]\label{exam:bessel} Given $\alpha > 0$, take $L u = -4(x u')' + \alpha^2 x^{-1} u$ on $(0,\infty)$. Since a~fundamental system of solutions of $L u = 0$ is given by $u(x) = x^{\pm \alpha/2}$, the endpoint $x=0$ is LP for $\alpha\geq 1$ and LC otherwise; the endpoint $x=\infty$ is LP in both cases. Fixing the LP case at $x=0$, we restrict ourselves to the case $\alpha \geq 1$.

For a given $\Im z >0$ the solutions $u_a$ ($u_b$) of $(L -z) u= 0$ being $L^2$ at $0$ ($\infty$) are spanned by \cite[equations~(9.1.7-9) and (9.2.5-6)]{MR0167642}
\begin{gather*}
u_a(x) = J_\alpha\big(\sqrt{x z}\big),\qquad u_b(x) = J_\alpha\big(\sqrt{x z}\big) + i Y_\alpha\big(\sqrt{x z}\big).
\end{gather*}
Thus, Theorem~\ref{thm:weyl} applies: $L$ is essentially selfadjoint on $C^\infty_0(0,\infty)$, the resolvent of its unique selfadjoint extension is represented, for $\Im z>0$, by the Green's kernel
\begin{gather*}
G_z(x,y) = \frac{i \pi}{4} \begin{cases}
J_\alpha(\sqrt{x z})\big(J_\alpha\big(\sqrt{\smash[b]{y z}}\big) + i Y_\alpha\big(\sqrt{\smash[b]{y z}}\big)\big), & x > y, \\
J_\alpha(\sqrt{\smash[b]{y z}})\big(J_\alpha\big(\sqrt{x z}\big) + i Y_\alpha\big(\sqrt{x z}\big)\big), & \text{otherwise}.
\end{cases}
\end{gather*}
For $\lambda >0$ there is the limit
\begin{gather*}
\pi^{-1} \Im G_{\lambda+i0}(x,y) = K_\lambda(x,y) = \frac{1}{4}J_\alpha\big(\sqrt{x \lambda}\big) J_\alpha\big(\sqrt{\smash[b]{y \lambda}}\big),
\end{gather*}
for $\lambda\leq 0$ the limit is zero; both limits are locally uniform in $x,y \in \R$. Hence, Theorem~\ref{thm:kernel} applies: $\sigma(L)=\sigma_\text{ac}(L)=[0,\infty)$ and (\ref{eq:projkern}) holds for each Borel set $\Delta \subset \R$. Given $0\leq s<\infty$, we may estimate for the specif\/ic choice $\Delta = (-\infty,1)$ that
\begin{gather*}
\left(\int_{0}^{s} \int_{0}^{s} \left(\int_\Delta |K_\lambda(x,y)| d\lambda\right)^2 dx dy\right)^{1/2}
\leq \frac14 \int_0^s \int_{0}^{1} J_\alpha\big(\sqrt{x \lambda}\big)^2 d\lambda dx =\tau_{\alpha}(s).
\end{gather*}
Therefore, Theorem~\ref{thm:kernel} yields that $\1_{(-\infty,1)}(L)\1_{(0,s)}$ is Hilbert--Schmidt with the \emph{Bessel kernel}
\begin{gather*}
\int_{-\infty}^{1} K_\lambda(x,y) d\lambda = \frac14 \int_0^1 J_\alpha\big(\sqrt{x \lambda}\big) J_\alpha\big(\sqrt{y \lambda}\big) d\lambda \\
\qquad{} = \frac{J_\alpha(\sqrt{x}) \sqrt{\smash[b]{y}} J_\alpha'(\sqrt{\smash[b]{y}}) - \sqrt{x} J_\alpha'(\sqrt{x}) J_\alpha(\sqrt{\smash[b]{y}})}{2(x-y)},
\end{gather*}
restricted to $x,y \in (0,s)$. Here, the last equality is obtained from a Christof\/fel--Darboux type of argument: First, we use the underlying dif\/ferential equation,
\begin{gather*}
x J_\alpha\big(\sqrt{x \lambda}\big) = - 4 \frac{d}{d\lambda}\left(\lambda \frac{d}{d\lambda} J_\alpha\big(\sqrt{x \lambda}\big) \right) + \alpha^2\lambda^{-1} J_\alpha\big(\sqrt{x \lambda}\big),
\end{gather*}
and partial integration to obtain
\begin{gather*}
\frac{x}4 \int_0^1 J_\alpha\big(\sqrt{x \lambda}\big) J_\alpha\big(\sqrt{y \lambda}\big) d\lambda \\
\qquad{} = - \int_0^1 \frac{d}{d\lambda}\left(\lambda \frac{d}{d\lambda} J_\alpha\big(\sqrt{x \lambda}\big) \right) J_\alpha\big(\sqrt{y \lambda}\big) d\lambda + \frac{\alpha^2}{4} \int_0^1 \lambda^{-1} J_\alpha\big(\sqrt{x \lambda}\big) J_\alpha\big(\sqrt{y \lambda}\big) d\lambda \\
\qquad{}= -\frac12 \sqrt{x} J'_\alpha\big(\sqrt{x}\big) J_\alpha\big(\sqrt{\smash[b]{y}}\big) \\
\qquad\quad{}+ \int_0^1 \lambda \left(\frac{d}{d\lambda} J_\alpha\big(\sqrt{x \lambda}\big)\right) \left( \frac{d}{d\lambda} J_\alpha\big(\sqrt{y \lambda}\big)\right) d\lambda + \frac{\alpha^2}{4} \int_0^1 \lambda^{-1} J_\alpha\big(\sqrt{x \lambda}\big) J_\alpha\big(\sqrt{y \lambda}\big) d\lambda.
\end{gather*}
Next, we exchange the roles of $x$ and $y$ and substract to get the assertion. Since the resulting kernel is continuous, Theorem~\ref{thm:kernel} gives that $\1_{(-\infty,1)}(L)\1_{(0,s)}$ is a trace class operator with trace
\begin{gather*}
\tr \1_{(-\infty,1)}(L)\1_{(0,s)} = \tau_\alpha(s) \to \infty, \qquad s\to \infty.
\end{gather*}

To summarize, we have thus obtained the following lemma.
\begin{Lemma}\label{lem:bessel} Given $\alpha\geq 1$, the differential operator $L u = -4( x u')' + \alpha^2 x^{-1}u$ is essentially selfadjoint on $C^\infty_0(0,\infty)$. The spectrum of its unique selfadjoint extension is
\begin{gather*}
\sigma(L)=\sigma_{\rm ac}(L)=[0,\infty).
\end{gather*}
Given $0\leq s<\infty$, the operator $\1_{(-\infty,1)}(L) \1_{(0,s)}$ is trace class with kernel
\begin{gather}\label{eq:bessel}
K_\text{\rm Bessel}^{(\alpha)}(x,y) = \frac14 \!\int_0^1\!\! J_\alpha\big(\sqrt{x \lambda}\big) J_\alpha\big(\sqrt{y \lambda}\big) d\lambda
= \frac{J_\alpha(\sqrt{x}) \sqrt{\smash[b]{y}} J_\alpha'(\sqrt{\smash[b]{y}}) - \sqrt{x} J_\alpha'(\sqrt{x}) J_\alpha(\sqrt{\smash[b]{y}})}{2(x-y)}.\!\!\!\!\!
\end{gather}
\end{Lemma}

\begin{Remark}\label{rem:bessel} Lemma~\ref{lem:bessel} extends to $0\leq \alpha < 1$ if we choose the particular self\-adjoint realization of~$L$ that is def\/ined by the boundary condition~(\ref{eq:besselRB}), cf.~\cite[Example~10.5.12]{MR2153611}.
\end{Remark}
\end{Example}

\pdfbookmark[1]{References}{ref}
\LastPageEnding

\end{document}